\documentclass[12pt]{JHEP3}

\let\ifpdf\relax
\usepackage{ifpdf}
\usepackage{graphics}
\usepackage{graphicx}
\usepackage{amssymb}
\usepackage{amsmath}
\usepackage{slashed}
\usepackage{cite}
\usepackage{epsfig}
\usepackage{datetime}

\newcommand{\be}{\begin{equation}}
\newcommand{\ee}{\end{equation}}
\newcommand{\bal}{\begin{align}}
\newcommand{\bea}{\begin{eqnarray}}
\newcommand{\eea}{\end{eqnarray}}
\newcommand{\half}{\frac{1}{2}}
\newcommand{\bi}{\begin{itemize}}
\newcommand{\ei}{\end{itemize}}

\def\p{\partial}
\def\a{\alpha}
\def\b{\beta}

\def\r{\rightarrow}

\def\l{\lambda}
\def\e{{\epsilon}}
\def\tphi{\tilde{\phi}}
\def\O{\mathcal{O}}
\def\D{\mathcal{D}}

\def\half{\frac{1}{2}}

\def\p{\partial}

\def\eps{{\epsilon}}

\def\half{\frac{1}{2}}
\def\tphi{{\tilde\phi}}

\def\tphi{{\tilde \phi}}

\def\CO{{\cal O}}
\def\ZZ{{\mathbb Z}}

\title{Holographic interpretations of the renormalization group}

\author{Vijay Balasubramanian${}^{a,b}$, Monica Guica${}^{a}$, and Albion Lawrence${}^{c}$ \\
{\small  ${}^a$ David Rittenhouse Laboratory, University of Pennsylvania, \\ \ \  Philadelphia, PA 19104, USA\\
${}^b$ Laboratoire de Physique Th\'{e}orique, \'{E}cole Normale Sup\'{e}rieure, \\  \ \ 75005 Paris, France \\
${}^c$ Martin Fisher School of Physics, Brandeis University, \\ \ \ Waltham, MA 02454, USA}}

\preprint{
NSF-KITP-12-195 \\
 BRX-TH-662}

\abstract{
In semiclassical holographic duality, the running couplings of a field theory are conventionally identified with the classical solutions of field equations in the dual gravitational theory.  However, this identification is unclear when the bulk fields fluctuate.   Recent work has used a Wilsonian framework to propose an alternative identification of the running couplings in terms of non-fluctuating data; in the classical limit, these new couplings do not satisfy the bulk equations of motion.   We study renormalization scheme dependence in the latter formalism, and show that a scheme exists in which couplings to single trace operators realize particular solutions to the bulk equations of motion, in the semiclassical limit. This occurs for operators with dimension $\Delta \notin \frac{d}{2} + \ZZ$, for sufficiently low momenta. We then clarify the relation between the saddle point approximation to the Wilsonian effective action ($S_W$) and boundary conditions at a cutoff surface in AdS space.   In particular, we interpret non-local multi-trace operators in $S_W$ as arising in Lorentzian AdS space from the temporary passage of excitations through the UV region that has been integrated out.   Coarse-graining these operators makes the action effectively local.
}

\begin{document}

\section{Introduction}

The holographic renormalization group, as conventionally understood, relates the radial flow of classical solutions in asymptotically anti-de Sitter (AdS) spacetimes to the renormalization group (RG) flow of dual large-N gauge theories.  At its core, this relation arises from two simple facts: (a) the boundary values of bulk fields are the couplings of the dual field theory \cite{Witten:1998qj,Gubser:1998bc}, and (b) motion in the radial direction is related to scaling in the field theory \cite{Maldacena:1997re,Susskind:1998dq}.  The latter scaling relation makes it natural to suppose that the values of bulk fields at fixed radial positions should describe the running couplings at the associated scale in the dual field theory.   Taking this view, the bulk equations of motion were related to RG equations in the dual field theory \cite{Akhmedov:1998vf,deBoer:1999xf}, and a proposal was made for a bulk description of Wilsonian renormalization of the field theory \cite{Balasubramanian:1999jd}.   In the latter  approach, the path integral over the ``ultraviolet" (UV) region between a specified radial cutoff and the boundary at infinity was treated as a functional of  field values at the cutoff, and encoded the interactions generated by renormalization.  These approaches were supported by the finding that  radial flows in many classical solutions qualitatively reproduce the expected properties of highly nontrivial RG flows, at least at the fixed points (see for example \cite{Porrati:1999ew}), e.g., in the gravitational realization of chiral symmetry breaking \cite{Klebanov:2000hb}.     

Despite these successes, one should be uncomfortable with interpreting the values of fields at finite positions in the bulk as running couplings in a field theory.   The bulk path integral is over all values of fields at finite positions, and is not defined with Dirichlet boundary conditions at a finite radius.  In a quantum theory of gravity, it is not even clear that {\it any} sensible definition can be given of a path integral with Dirichlet conditions at finite radius.\footnote{Classical Dirichlet conditions at finite radius are discussed in \cite{Brattan:2011my,Marolf:2012dr}.}   Indeed, the standard dual interpretation of gravity in AdS space cut off at finite radius is as a conformal field theory coupled to lower-dimensional gravity \cite{Randall:1999ee,Randall:1999vf,Verlinde:1999fy}, with the couplings  appearing as functions of fluctuating moduli.   Thus, the only hope for a sensible relation between running couplings and the  values of fields  in AdS is in terms of saddle point approximations.   However, there can be multiple saddle points with the same asymptotic boundary conditions.   In this circumstance, which saddle point, or what sum over saddle points, describes the running coupling?  At any rate, it is desirable to have a fully quantum mechanical understanding of the map between quantities in asymptotically AdS spacetimes and the running couplings of the dual theory, which can be {\it approximated} using large $N$ saddle points.

Recently, Refs. \cite{Heemskerk:2010hk,Faulkner:2010jy} presented a revised approach to the holographic Wilsonian renormalization group in which the full path integral in the bulk spacetime was taken seriously.  We will describe this in more detail below, but the basic idea is as follows.  The path integral in the bulk spacetime is broken up at a fixed radial position $\ell$, identified via scale-radius duality with the running cutoff.  The path integral over the fields in the region between this cutoff and spatial infinity is roughly dual to an integral over the ultraviolet modes of the field theory, as a functional $\Psi_{UV}(\phi_{\ell})$ of the fields $\phi_{\ell}$ at the cutoff.  $\Psi_{UV}$ is defined by the boundary conditions on the fields at spatial infinity, which are dual to couplings in the bare theory following \cite{Witten:1998qj,Gubser:1998bc}.  As in \cite{Balasubramanian:1999jd}, the path integral $\Psi_{IR}(\phi_{\ell})$ over the remainder of the spacetime (with some regularity conditions in the interior) is {\it assumed}\ to be the generating function of vacuum correlators of single-trace operators in the cutoff theory, with $\phi_{\ell}$ identified as sources for these operators.  To complete the full path integral, one then integrates over the values of $\phi_{\ell}$.  The result is identified with a Wilsonian action for the cutoff theory, arrived at by integrating out ultraviolet degrees of freedom.  The couplings are derived from the parameters defining the functional $\Psi_{UV}$, and depend on the bare couplings and on $\ell$. Refs \cite{Heemskerk:2010hk,Faulkner:2010jy}\ then showed that the running of these couplings with $\ell$ corresponds to a set of renormalization group equations. Notably, the single-trace couplings as functions of $\ell$ do not satisfy the spacetime equations of motion for the dual scalar field and multiple-trace couplings are induced at leading order in the $1/N$ expansion.

In this paper, we wish to better understand the relationship between these proposals, and then to develop the proposal of \cite{Heemskerk:2010hk,Faulkner:2010jy}\ further.  In \S2\ we review and compare the work of \cite{deBoer:1999xf,Balasubramanian:1999jd,Heemskerk:2010hk,Faulkner:2010jy}. In \S3\ we investigate the proposed identification of $\Psi_{IR}$ \cite{Heemskerk:2010hk,Faulkner:2010jy} with the generating function of correlation functions in the cutoff theory, for scalar fields in a fixed AdS background.  We find that the proposal uses a specific scheme, in which the correlators essentially include an infinite set of contact terms.  We will provide criteria for alternative schemes in defining this generating function, and define a sensible ``minimal subtraction" scheme.   Furthermore, we will find that for spacetime dependent couplings/sources with momenta  sufficiently below the cutoff, a scheme exists for most operators in which all of the contact terms are removed and the couplings are a particular solution to the equations of motion. In \S4\ we recall that in \cite{Heemskerk:2010hk,Faulkner:2010jy}, nonvanishing, nonlocal multiple-trace operators are induced at the cutoff even for an unperturbed CFT.   We describe $\Psi_{UV}$ as defining boundary conditions for the path integral over the IR region; for classical solutions, these boundary conditions encode the propagation of signals through the UV region.  We thus interpret the multiple-trace operators in the dual CFT as transporting excitations of the theory into and out of the UV region.  We discuss the relationship of this phenomenon to recent work on quantum entanglement of interacting theories in momentum space \cite{Balasubramanian:2011wt}, and argue that upon appropriately coarse-graining the observables of the theory, the multi-trace operators can nonetheless describe deformations of a local Hamiltonian for the infrared theory. In \S5\ we discuss some directions for future work.

While this work was being completed, the complementary paper \cite{Dong:2012af}\ appeared with a counterterm prescription at the running cutoff that is similar to ours.  Those authors also point out that the multi-trace couplings run even for unperturbed conformal field theories. That work focuses on the structure of constant perturbations away from a fixed point, and on the role of the nondynamical scale factor.  Our work focuses on the momentum dependence of the Wilsonian action and correlation functions, for scalars in a fixed AdS background. 

\section{Approaches to the Holographic Renormalization Group}


To set the stage for our results, we begin by reviewing prior approaches to the holographic renormalization group  
\cite{deBoer:1999xf, deBoer:2000cz,Balasubramanian:1999jd}, and  the recent proposal of \cite{Heemskerk:2010hk,Faulkner:2010jy}.   To establish our notation, we summarize the standard formalism for regularization and renormalization of the bare holographic theory  \cite{Witten:1998qj,Gubser:1998bc,Henningson:1998gx,Balasubramanian:1999re,deHaro:2000xn,Bianchi:2001de,Bianchi:2001kw} (reviewed in \cite{Skenderis:2002wp}).

We will focus on the dynamics of scalar fields in a fixed anti-de Sitter (AdS) background, using the Poincar\'e patch metric:
\be
	ds^2 = R^2 \, \frac{dz^2 + \eta_{\mu\nu} dx^{\mu}dx^{\nu}}{z^2} \, \label{poincm}
\ee
where $\eta_{\mu\nu}$ is the $d$-dimensional Minkowski metric,  $R$ is the AdS radius of curvature,  the space-time boundary is at $z \to 0$, and  the Poincar\'e horizon is at $z \to \infty$.   The scalar field action is
\be
\label{eq:scact}
	S_{bulk} = - \half N^2 
\int d^{d+1} x \, \sqrt{g} \, \left[ (\p \phi)^2 + m^2 \phi^2 + V(\phi) \right] \, .
\ee
Here $N $ 
 is the rank of the gauge group for the cases with gauge theory duals, and $V(\phi)$ contains cubic and higher interactions. The field $\phi$ is dual to a single-trace field theory operator $\O$ with scaling dimension:
\be
 \Delta = \frac{d}{2} + \sqrt{\frac{d^2}{4} + (m R)^2} \equiv \frac{d}{2} + \nu
\ee
For simplicity, we will consider the case  $\nu \notin\ZZ$.

Consider the case that $V = 0$. As $z \to 0$, solutions to the equations of motion take the general form
\be\label{eq:asymptotics}
	\phi \sim z^{d - \Delta} \left(\alpha({\vec x}) + \a_{(2)}({\vec x}) z^2 + \ldots\right) + z^{\Delta} \left(\beta({\vec x}) + \b_{(2)}({\vec x}) z^2 + \ldots \right) \, ,
\ee
where the coefficients $\a_{(k)}$, $\b_{(k)}$ are determined by $\alpha,\beta$ respectively.  For general $\Delta$, the piece proportional to $\alpha(x)$ grows fastest near the boundary at $z \to 0$.  For $\nu > 1$, this piece is non-normalizable, and hence cannot fluctuate.  Thus good boundary conditions for $\phi$ in this mass range will have $\alpha$ fixed.\footnote{For $\nu < 1$ there are additional consistent choices for the boundary conditions \cite{Balasubramanian:1998sn,Klebanov:1999tb}, which describe multiple-trace deformations of an operator with dimension $d - \Delta$ \cite{Aharony:2001pa,Witten:2001ua,Berkooz:2002ug,Sever:2002fk}.}

In these theories the action has divergences, due to boundary contributions which are functions of $\alpha$.  To regulate these, one first cuts off AdS space at a large radius $z = \eps$ and imposes  Dirchlet conditions $\phi(\eps,x) = \eps^{d - \Delta} \a(x)$.  One then adds a ``boundary term" to the spacetime action which is a covariant functional of the scalar, the induced metric $h_{\mu\nu}$ at $z = \eps$, and their derivatives along the $z = \eps$ surface:
$	
	S_{ct}^\e = \int d^d x \, \sqrt{h} \,  {\cal L}(\phi,\p_{\mu_1}\ldots \p_{\mu_{n}}\phi, h_{\mu\nu}) .
$
  This counterterm action can always be chosen to render the action finite on solutions to the equations of motion  \cite{Henningson:1998gx,Balasubramanian:1999re,deHaro:2000xn,Bianchi:2001de,Bianchi:2001kw,Skenderis:2002wp}.  One then takes the limit $\eps \to 0$, keeping $\alpha(x)$ fixed.  Given this holographically renormalized, finite action the standard correspondence relates the  path integral with Dirichlet boundary conditions (fixed $\alpha(x)$) to the generating function of correlation functions in the dual field theory
\be
\label{adscftdef}
\mathcal{Z}[\a] =	\int_{\phi \stackrel{z\to 0}{\to} \alpha(x) \, z^{d-\Delta} }D\phi \, e^{i (S + S_{ct}^\e)} = \Big\langle \exp\left( i N^2 \int d^d x \, \alpha(x) \CO(x)\right)\Big\rangle
\ee
so that $\alpha(x)$ is the field theory coupling. 

In the large $N$ limit, the correlation functions of local operators in the dual field theory are computed by finding  solutions to the bulk classical equations of motion which are regular in the interior and which have Dirichlet boundary conditions as $z \to 0$.   Correlation functions are obtained by varying the action evaluated on these solutions as a function of the Dirichlet conditions.  By doing this, one is varying the classical action over different classical solutions.  

\subsection{Running couplings from classical solutions?}

In the AdS-CFT duality, the scaling symmetry of the CFT is dual to the isometry $z \to \lambda z\ ,  \  x^{\mu} \to \lambda x^{\mu}$. This and other arguments \cite{Susskind:1998dq,Peet:1998wn}\ motivate identifying the radial position $z$ with an energy scale $\Lambda \sim 1/z$. In the classical  limit, it is natural to associate solutions $\phi(z,x)$ with running (spacetime-dependent) field theory couplings. 


This association was formalized in \cite{deBoer:1999xf, deBoer:2000cz,Martelli:2002sp,Papadimitriou:2004ap}.
Consider the classical bulk action $S(\phi_\Lambda, g_{\Lambda,\mu\nu})$ evaluated in the region $z > \Lambda^{-1}$, with boundary conditions $\phi(z = \Lambda^{-1},x) = \phi_{\Lambda}$, $g_{\mu\nu}(z = \Lambda^{-1},x) = g_{\Lambda,\mu\nu}$.  
This solves the Hamilton-Jacobi equations for flow in the radial direction.  In Euclidean space, regularity as $z \to \infty$ uniquely selects a solution with these boundary conditions. Refs. \cite{deBoer:1999xf, deBoer:2000cz} then set
\be
	S = S_{ct} + \Gamma
\ee
where the counterterm action $S_{ct}$ can be written as a local functional of $\phi_{\Lambda},g_{\Lambda}$ and removes the divergent parts of the bare action as $\Lambda \to \infty$, as discussed above.  $e^{-\Gamma}$ is the generating function of vacuum correlators of the operator $\O$ in the CFT cut off at the scale $\Lambda$, in the large $N$ limit. Refs. \cite{deBoer:1999xf, deBoer:2000cz}\ argue that the Hamilton-Jacobi equations can be broken up into equations for the divergent counterterms, and Callan-Symanzik equations for $\Gamma$.

%

Taken together, the Hamilton-Jacobi equations and Hamilton's equations are equivalent to the second-order spacetime equations of motion.  If we perturb the CFT by specifying $\phi(\eps,x) = \eps^{d-\Delta}\alpha(x)$, then $\phi(z,x)$ are interpreted as the running couplings at scale $\Lambda = z^{-1}$. Hamilton's equations are mapped to the renormalization group flow equations via the association
\be\label{eq:rgflow}
	 z \frac{d}{dz} \phi = \frac{\delta {\cal H}}{\delta \pi_{\phi}} \Big|_{ \pi_{\phi} = \frac{\delta S}{\delta\phi}}  ~~~~~\longleftrightarrow~~~~~~ \Lambda \frac{d}{d\Lambda} g = \beta(g).
\ee
Similarly the radial flow of the the metric is related to the flow of couplings for a stress tensor deformation of the field theory.\footnote{Note that the data specifying a general solution to the second-order equations of motion is encoded in the choice of solution to the Hamilton-Jacobi equations \cite{Lawrence:2006ze}, as well as the initial conditions needed to solve Hamilton's equations.}   

There are are a number of confusing aspects in this formalism, discussed variously in \cite{Lawrence:2006ze,Heemskerk:2010hk}, especially when using it at finite cutoff and in Lorentzian signature \cite{Lawrence:2006ze}.  In this case, the equation of motion for $\phi$ has two independent solutions which are nonsingular as $z\to \infty$, differentiated by their behaviour near the boundary as in (\ref{eq:asymptotics}).
 Recall that in the general solution (\ref{eq:asymptotics}) $\alpha$ is dual to the field theory coupling, while $\beta$ is related  to the state of the theory \cite{Balasubramanian:1998sn,Balasubramanian:1998de}.    As $z \to 0$, we can unambiguously identify the field $\phi$ with the coupling  $\alpha$  in the standard quantization (for $\Delta > d/2$), because this component of the solution grows fastest. In this limit, the formalism has been successfully adapted to deal with the different states that the field theory could find itself in \cite{Lawrence:2006ze, Papadimitriou:2004ap}. Nevertheless, 
 at finite $z$, the value of $\phi$ is controlled by both $\alpha$ and $\beta$.  While we might expect the running of the couplings to be affected by the quantum state of the UV degrees of freedom (and thus to depend on both $\alpha$ and $\beta$), it is not clear how to unpack this information from $\phi_{\Lambda}$.\footnote{This issue has also been discussed in \cite{Porrati:1999ew,Muck:2010uy}.}

Another basic issue is that this is at best an approach suited to the semiclassical approximation.  As we discussed in the introduction, the identification of $\phi_{\Lambda}$ with the running coupling cannot  be precise in the quantum theory.

\subsection{Wilsonian holographic renormalization I}

A first attempt at a Wilsonian approach was given by  \cite{Balasubramanian:1999jd}.  The basic ingredients are: (a) the  duality map equating the supergravity partition function with Dirichlet conditions as $z \to 0$, for fields defined over the entire spacetime, with the partition function of the full 
 perturbed CFT, and (b) the assumption that the field values at a given radial position are equivalent to the running couplings at the dual CFT scale. The latter assumption makes sense only in the classical limit, so we will review this formalism strictly in that limit.

Consider a solution to the supergravity equations with fixed $\alpha(x)$ as in (\ref{eq:asymptotics}).  Regularity in the interior -- that is, the assumption that the theory is in the vacuum state -- will then uniquely specify a solution in the interior \cite{Witten:1998qj}.\footnote{This is true in the free theory; in an interacting theory, one may have to cope with multiple saddle points.}  We are interested in studying the theory at a cutoff scale $\Lambda = \ell^{-1}$, after integrating out the ultraviolet degrees of freedom.  The basic formulation of  \cite{Balasubramanian:1999jd} can be mapped out as follows:

\begin{itemize}
\item Let $ \phi^{(s)}_{\ell}(x)  \equiv \phi (z=\ell, x) $ be the unique classical solution that  satisfies the above requirements.  This will be the running coupling, and is a function of $\alpha$, the boundary condition at the AdS boundary.  

\item In the classical limit, the path integral over fields in the IR region is

\be
	\Psi_{IR}(\phi_{\ell}^{(s)}) = \int_{z > \ell} D\phi \, e^{- S(\phi)} ~~~\stackrel{N\to\infty}{\longrightarrow}~~~ e^{-S^{cl}_{IR}(\phi_{\ell}^{(s)})} \equiv \Big\langle e^{-N^2\int d^d x \, \Lambda^{d-\Delta} \, \phi^{(s)}_{\ell}\O} \Big\rangle_{CFT, \Lambda}
\ee
Here we have borrowed the symbol $\Psi_{IR}$ from \cite{Heemskerk:2010hk}.  The final term in the right hand side represents the generating function of correlation functions in the CFT cut off at the scale $\Lambda$.  The classical action is evaluated on the solution that approaches $\phi_\ell^{(s)}$ at $z=\ell$ and is regular in the interior.
\item One can further define\footnote{A similar definition was given in \cite{Akhmedov:2002gq}.}

\be
	\Psi_{UV}(\phi_{\ell}^{(s)}) = \int_{z < \ell} D\phi \, e^{- S(\phi)}~~~ \stackrel{N\to\infty}{\longrightarrow}~~~ 	
 e^{-S^{cl}_{UV}( \phi_{\ell}^{(s)})} 
\ee
Again, we have borrowed the symbol $\Psi_{UV}$ from  \cite{Heemskerk:2010hk}.  The meaning of the above manipulations is the following: the classical action in the region $0< z <\ell$ is evaluated on the  solution defined by fixed $\alpha$ as $z\to 0$ and by $\phi_{\ell}^{(s)}(\a)$ at $z=\ell$. One then inverts the relationship between $\a$ and $\phi_{\ell}^{(s)}$ to trade all the $\a$-dependence for dependence on $\phi_{\ell}^{(s)}$.
 
\item The candidate Wilsonian action, as a function of $\phi_{\ell}^{(s)}$, is identified as:

\be
	\mathcal{Z}_{Wilson}[\phi_{\ell}^{(s)}] = \Psi_{UV,cl}\left(\phi_{\ell}^{(s)}(\alpha)\right)\Psi_{IR,cl}\left(\phi_{\ell}^{(s)}(\alpha)\right)
		\equiv  \Big\langle e^{-N^2 \int d^d x \, \Lambda^{d-\Delta} \phi_{\ell}^{(s)}\O} \Big\rangle_{\Lambda}\label{eq:bkwilson}
\ee
In this formula, the final expectation value is taken in the full theory corresponding to a perturbed CFT with coupling $\alpha$ in the UV, with the UV modes at scales higher than $\Lambda$ integrated over.  $\phi_{\ell}^{(s)}$ are the running couplings at the scale $\Lambda = \ell^{-1}$, determined by the couplings $\alpha$ in the bare theory.  

\item If one wishes to compute correlation functions of operators defined at the scale $\Lambda$, one differentiates $\mathcal{Z}_{Wilson}[\phi_{\ell}^{(s)}]$ with respect to $\phi_{\ell}^{(s)}$. One can use the classical map between $\alpha$ and $\phi_{\ell}^{(s)}$ to relate these correlation functions to those in the un-flowed theory. 
 This map is the bulk-boundary propagator.
\end{itemize}

As in \S2.1\ above, the identification of $\phi_{\ell}^{(s)}$ becomes confusing if one continues to Lorentzian signature, in which case it is also determined by the state.  More generally, it remains unclear what the correct bulk dual of the running couplings is in the quantum theory.  Finally, because the equations of motion are second order, $\phi_{\ell}^{(s)}$ is determined by $\alpha$ together with the regularity conditions in the IR region.  This is in tension with the idea of Wilsonian renormalization, where the effective action at a certain cutoff depends only on the UV degrees of freedom that were integrated out, and not on the physics below the cutoff scale. 

\subsection{Wilsonian holographic renormalization II}

Recent work \cite{Heemskerk:2010hk,Faulkner:2010jy}\ provides a different prescription for the running couplings that takes the  full path integral seriously.  In this work, the integrand of the path integral is broken up as in (\ref{eq:bkwilson}), but $\phi_{\ell}$ (which is no longer required to solve the equations of motion) is integrated over. 
For specificity, consider a  single scalar field propagating on a fixed AdS background.  The full Lorentzian path integral is written as:
\be
	\mathcal{Z}[\a] = \int {\cal D} \phi_{(z<\ell)} \,  {\cal D}\phi_\ell \, {\cal D}\phi_{(z > \ell)} \, e^{i  S_{sugra}}
\ee
where $\phi_{\ell} = \phi(z = \ell)$, and $\ell \equiv \Lambda^{-1}$ will become the spatial cutoff scale.  
 
We define:
\begin{eqnarray}
	\Psi_{UV}(\alpha,\phi_{\ell}) & = & \int_{\phi(\ell,x) = \phi_{\ell}}^{\phi(\eps,x) \sim \eps^{d-\Delta}\alpha}
	\D\phi_{(z \leq \ell)}\, e^{i S_{sugra}} \nonumber\\ &&  \\
	\Psi_{IR}(\phi_{\ell}) & = & \int_{\phi(\ell,x) = \phi_{\ell}(x)} {\cal D}\phi_{(z \geq \ell)} \, e^{i  S_{sugra}} 	
\end{eqnarray}
In the first line, $\eps$ is understood to approach zero, after adding counterterms at $z = \eps$.  Next, in analogy to \cite{Balasubramanian:1999jd}, we make the assumption that

\be\label{eq:hpansatz}
	\Psi_{IR}(\phi_{\ell}) =
		\Big\langle e^{i N^2 \int d^d x \ell^{\Delta-d} \phi_{\Lambda}\O} \Big\rangle_{CFT, \Lambda = \ell^{-1}}
\ee
The right hand side is the generating function of correlation functions in the cutoff CFT, as in \S2.2.  This is a supposition (though a plausible one), which is at the heart of the proposal of \cite{Heemskerk:2010hk,Faulkner:2010jy}. We will investigate it further in \S3. Note that these definitions evade the question of whether the quantum path integrals with such boundary conditions make sense in a theory of quantum gravity.  We will not address this issue here.

Note that we have not been careful about specifying the state here.  We will take the theory to be in the vacuum, as does Ref. \cite{Heemskerk:2010hk}.  For scalars with $\nu < 1$, we could instead consider the alternate quantization \cite{Klebanov:1999tb}, or double-trace perturbations of that quantization \cite{Aharony:2001dp,Aharony:2001pa,Witten:2001ua}. We will focus on the standard quantization, with Dirichlet boundary condition at $z = \eps$, in this paper.

The full path integral can now be rewritten as:

\be\label{eq:segmented}
	\mathcal{Z}[\a] = \int \D\phi_{\ell} \, \Psi_{UV}(\alpha,\phi_{\ell}) \, \Psi_{IR}(\phi_{\ell}) 
\ee
Once $\Psi_{UV}$ has been calculated, this gives a specific identification of the running couplings in terms of parameters of the functional $\Psi_{UV}$.  To see this, suppose (following \cite{Heemskerk:2010hk}) that\footnote{$\Psi_{UV}(\phi_\ell)$ will always be quadratic in $\phi_\ell$ for a free field.} 
\be
	\Psi_{UV}(\phi_{\ell}) = \exp\left[ i N^2 \int d^d x \, \frac{1}{2h\ell^{2(d-\Delta)}} (\phi_\ell(x) - \ell^{d-\Delta} \l)^2\right]
\ee
where $h,\l$ are functions of $\ell,\alpha$, which for simplicity we assume to be constant. Inserting the CFT dual of $\Psi_{IR}$, i.e. (\ref{eq:hpansatz}) into (\ref{eq:segmented}), the full path integral can be written as
\be
	\mathcal{Z} = \Big\langle e^{i N^2 \int d^d x \left( \l \O + \frac{h}{2} \O^2\right)} \Big\rangle_{CFT, \ell^{-1}}
\ee
This is defined as the Wilsonian action and $\l,h$ are running couplings for the single- and double-trace operators. In effect, (\ref{eq:segmented}) implies that the Wilsonian action arises as a Hubbard-Stratonovich transformation of the generating function of correlators of single-trace operators. 

More generally, $\Psi_{UV}$ will be some more complicated functional of $\phi_{\ell}$,
and we will find:
\be
\mathcal{Z} = \Big\langle e^{i N^2 \int d^d x \sum_n \l_n \O^n } \Big\rangle_{CFT,\ell^{-1}} 
\ee
where $\l_n(\ell)$ are functions of the UV data and of the cutoff $\ell$, and can be multilocal functions of the spacetime coordinates.  The Hamilton-Jacobi equations for $\Psi_{UV}$ then provide a set of renormalization group equations for the couplings $\l_n$ \cite{Heemskerk:2010hk,Faulkner:2010jy}. 

The single trace coupling $\l_1$ in the above expression turns out to have a simple interpretation in terms of parameters in $\Psi_{UV}$: it is the solution to the equation
 $\frac{\delta S_{UV}(\phi_{\ell})}{\delta \phi_{\ell}} = 0$. Indeed, taking $\lambda_1$ to solve this equation, $S_{UV}$ can be expanded in a power series in $(\phi_{\ell} - \l_1)$ with vanishing first-order term:
\be
	S_{UV} = N^2 \sum_{k=0}^{\infty} \int C_k (\phi_{\ell} - \l_1)^k\ , \ \ C_1 = 0
\ee
This expression is schematic; in practice, $C_k$ is a nonlocal kernel. To evaluate the integral (\ref{eq:segmented}) by the saddle point method, we solve
\be
	\frac{\delta \left(S_{UV} + N^2 \int d^d x \, \ell^{d - \Delta} \phi_{\ell}(x) \O(x)\right)}{\delta \phi_{\ell}} = 0
\ee
We can solve for $\phi_{\ell} - \l_1$ in a power series in $\O$, and write
\be
	\mathcal{Z} =  \Big\langle e^{i N^2 S_{UV}(\phi_{\ell}(\O)) i N^2 \int \ell^{d-\Delta} \phi_{\ell}(\O) \O} \Big\rangle\ .
\ee
The argument of the exponential can be expanded in $\O$, and one finds that the term linear in $\O$ is $\l_1\O$. Note that there is no reason for $\l_1(x,\ell)$ to be a solution to the equations of motion if we replace $\ell$ with $z$. Note also that there will be multiple-trace couplings to leading order in the $1/N$ expansion \cite{Heemskerk:2010hk,Faulkner:2010jy}.\footnote{This was also pointed out earlier in \cite{Akhmedov:1998vf,Akhmedov:2002gq}.}

\bigskip

\FIGURE[h]{
\includegraphics[scale=.5]{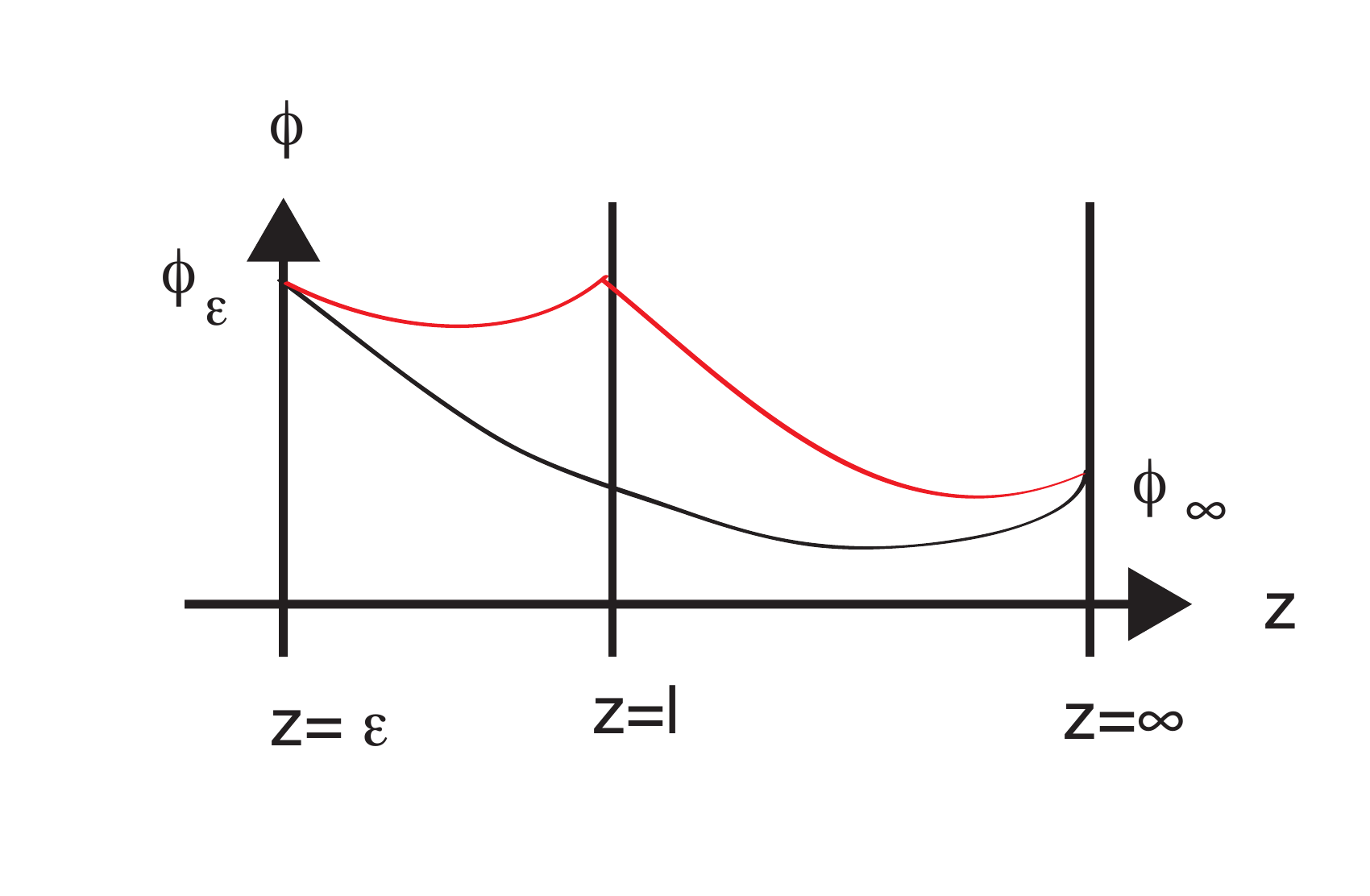}
\caption{Cartoon of the saddle point evaluation of the full path integral (\ref{eq:segmented}).  If one finds the saddle points for $\Psi_{UV}$ and $\Psi_{IR}$ separately, the saddle points will not join smoothly at $z=\ell$ for general $\phi_{\ell}$, as illustrated by the red line, which contains a kink at $z = \ell$. If one then minimizes $\Psi_{UV}\cdot \Psi_{IR}$ with respect to $\phi_{\ell}$, $\phi_{\ell}$ will take a value such that the saddle points for $\psi_{UV},\Psi_{IR}$ join smoothly, as illustrated by the black line.}
\label{kinks}
}

To  appreciate the similarities and differences of this work to prior work, it is useful to
understand the evaluation of the full path integral \eqref{eq:segmented} via  saddle points. In this approximation,
 $\Psi_{UV}(\alpha,\phi_{\ell})$  is determined by a solution to the classical equations of motion in the region $0 < z \leq \ell$, with boundary conditions $\phi(z,x) \to_{z\to 0} z^{\Delta-d}\alpha(x)$ and $\phi(\ell,x) = \phi_{\ell}$. Likewise, $\Psi_{IR}(\phi_{\ell})$ is determined by the classical solution  which is regular in the interior and  has boundary condition $\phi(\ell,x) = \phi_{\ell}(x)$.  For general $\phi_{\ell}$, the two solutions will not match together smoothly at $z = \ell$, as illustrated by the red line in Figure (\ref{kinks}).  Furthermore, the solution in the region $z > \ell$ depends on specifying the boundary condition (state) in the IR.

The final step in the saddle point evaluation of the path integral (\ref{eq:segmented}) is to vary the integrand over $\phi_{\ell}$ and look for the saddle point.  (This is not the same as the saddle point of $\psi_{UV} \langle e^{i N^2 \int \phi_{UV} \O} \rangle$ that determines the running Wilsonian coupling in the saddle point approximation).  This will be the smooth solution illustrated by the black line in Figure (\ref{kinks}).  This solution is determined by $\alpha$ and by the boundary conditions in the IR (eg regularity); it is the same solution that appears in every version of holographic renormalization.  Note that had we chosen a different boundary condition in the IR, corresponding to some nontrivial state, $\Psi_{IR}$ would be a different functional of $\phi_{\ell}$. Even if $\Psi_{UV}$ remains unchanged, the value of $\phi_{\ell}$ on the saddle point of the full path integral will be different, since it depends on $\Psi_{IR}$.  This is one indication that in a Wilsonian framework, the value $\phi_{\ell}$ on the saddle points of the full path integral is not a good candidate for the running coupling.  

The parameters defining $\Psi_{UV}$ are independent of $\Psi_{IR}$.  The running couplings $\l_k$ are determined by these parameters alone.  They are well defined even in the quantum theory, and independent of the state of the IR degrees of freedom.  Again, in contrast to prior work, the single trace couplings in \cite{Heemskerk:2010hk,Faulkner:2010jy}\ do not directly satisfy the spacetime equatons of motion as functions of $\ell = z$.  Furthermore, they include multiple-trace couplings at leading order in the $1/N$ expansion. We will revisit these observations below.

\section{Holographic RG and scheme dependence}

\def\Z{{\mathbb Z}}

In this section we explore renormalization scheme dependence in the formalism of \cite{Heemskerk:2010hk,Faulkner:2010jy}\ for a scalar field on a fixed background, with Lagrangian (\ref{eq:scact}).   We  find that taking the proposed identification (\ref{eq:hpansatz}) at face value implicitly uses a specific scheme, in which $\phi_{\ell}$ couples to a combination of $\O$ and other operators.  The renormalization scheme can be changed by adding counterterms to $S_{IR}$ and subtracting them from $S_{UV}$, such  that the full bulk action remains unchanged.
 
 We  first define a ``minimal" scheme, which adds a finite number of counterterms matching the prescription at the UV cutoff $z = \eps$, and compute the corresponding running couplings.  We  then show that for sufficiently low momenta, there is also a ``maximal'' scheme, in which the single-trace couplings are a particular, non-normalizable solution to the equations of motion.  This result establishes a connection with the previous view (summarized in $\S$2), that the radial flow of classical solutions in AdS space reproduces the renormalization group flow of couplings.   In \S3.3 we argue that this scheme can be extended to the interacting theory.  

%

\subsection{$\Psi_{UV}$ and $\Psi_{IR}$ for a free massive scalar in AdS}

We begin by explicitly computing $\Psi_{UV}$ and $\Psi_{IR}$ for a free scalar governed by the Lagrangian (\ref{eq:scact}) with $V = 0$.  
To leading order in $1/N$ (that is, ignoring the one-loop determinants), both are given by the exponential of the classical on-shell action 
\be
\Psi_{UV} = e^{i S_{UV}} \;,\;\;\;\;\;\; \Psi_{IR} = e^{i S_{IR}}
\ee
evaluated with appropriate boundary conditions at $z=\ell$ and  $z = \epsilon$ and properly renormalized. We only consider scalar fields $\phi$ dual to relevant operators with dimension $\Delta < d$.  For simplicity, we also assume that $\nu \equiv \Delta - \frac{d}{2} \notin \Z$.  We  work in Fourier space, with
\be
	\tphi(q,z) = \int\frac{d^d q}{(2\pi)^d}\, e^{i q\cdot x} \phi(z,x) \, ,
\ee
The general solution to the equations of motion is:

\be
	\tphi(z,q)= \left\{  \begin{array}{lll} \alpha_q\, \Gamma(1-\nu) \, (\half q)^\nu \, z^{\frac{d}{2}} I_{-\nu}(qz) + \beta_q\, \Gamma(1+\nu) \, (\half q)^{-\nu}\, z^{\frac{d}{2}} I_{\nu}(qz)& \hspace{1cm} &
		q^2 > 0 \\ & \\
	\alpha_q \, \Gamma(1-\nu) \,  (\half q)^\nu  \,z^{\frac{d}{2}} J_{-\nu}(qz) + \beta_q \, \Gamma(1+\nu) \, (\half q)^{-\nu}\, z^{\frac{d}{2}} J_{\nu}(qz)& \hspace{1cm} &
		q^2 < 0 \label{eq:eomsol} \end{array} \right. 
\ee
where $q^2 > 0$ corresponds to spacelike or Euclidean momenta, and $q^2 < 0$ to timelike ones.
Here $I_{\nu}, J_{\nu}$ are Bessel functions, defined for example in \cite{abramowitz1965handbook}. 
If we consider the solution to be defined over all of AdS spacetime, normalizability as $z \to \infty$ requires that $\b_q = - (q/2)^{2\nu} \frac{\Gamma(1-\nu)}{\Gamma(1+\nu)}\, \a_q$, for the spacelike case $q^2 >0$.


\subsubsection{Computing $\Psi_{UV}$ }

The path integral $\Psi_{UV}(\phi_{\ell}) = e^{i S_{UV}(\phi_{\ell})}$ is defined in \cite{Heemskerk:2010hk}\ by integrating the bulk fields in the region $\ell > z > \eps$. The UV theory is defined by the boundary conditions at $z = \eps$; these are Dirichlet if we consider the standard quantization.  At $z=\ell$ the boundary condition is simply that $\tilde \phi(z=\ell,q) = \tphi_\ell(q)$. The full path integral (\ref{eq:segmented}) should be identical to that used in the standard treatments of \cite{Witten:1998qj,Gubser:1998bc,Henningson:1998gx,deHaro:2000xn,Bianchi:2001de,Bianchi:2001kw,Skenderis:2002wp}.  This means that we should identify the terms which are divergent as $\eps \to 0$, add counterterms which subtract these divergences, and take the $\eps \to 0$ limit.

%

The bare on-shell action with Dirichlet boundary conditions at $z=\ell$ and $z = \e $ is
\be
S_{UV}^{\e} = \frac{N^2}{2} \int d^{d+1}x \sqrt{g}\, \phi (\Box - m^2 ) \phi + \frac{N^2}{2} \left. \int d^d x \sqrt{h}\, \phi \, z\p_z \phi \, \right|_{z=\e}^{z=\ell} 
\ee
The bulk part of the integral over the region $\ell<z<\e$ vanishes by the equations of motion. The $\eps\to 0$ limit is defined with a scaling $\tphi_{\eps}(q) = \tphi(z=\eps,q) = \eps^{\Delta_-} \a_q$, where $\a_q$ is held fixed and $\Delta_- = d - \Delta$. The coefficient $\a_q$ has the correct dimension to be identified with the coupling to an operator of dimension $\Delta$. With this scaling, the contribution from the boundary at $z = \eps$ includes a finite set of divergent terms, multiplied by positive powers of $q^2$.  These can be removed by adding  a counterterm action $S_{ct}^\e$ to $S_{UV}^\e$.  In $d = 4$, for $\Delta < d$ the counterterms are:

\be
S_{ct}^\e = \frac{N^2}{2} \int_{z=\e} d^4x \, \sqrt{h} \left[ \Delta_- \phi_{\eps}^2 + {(\nabla \phi_{\eps})^2 \over 2\nu - 2} \right] = 
 \frac{N^2}{2} \int \frac{d^4 q}{(2\pi)^4} \, \left(\frac{\Delta_-}{\eps^4} + \frac{q^2}{\eps^2(2\nu-2)}\right) \tphi_{\eps}^2
\ee
For $\nu \in \Z$, there are additional logarithmic counterterms, as well as finite terms one can add.  The latter  contribute contact terms in the correlation functions of $\O$; one may choose a scheme in which they are also subtracted. 

Adding $S_{ct}^\e$ and then taking $\eps \to 0$ with $\a_q$ fixed is the holographic renormalization procedure of \cite{Henningson:1998gx,deHaro:2000xn,Bianchi:2001de,Bianchi:2001kw,Skenderis:2002wp}.  Applying this procedure to $S_{UV}^\e$, we find the expression:

\begin{eqnarray}
S_{UV} & = & \frac{N^2}{2} \int \frac{d^d q}{(2\pi)^d}\left\{
	\left(\frac{2^{1-2\nu} \Gamma(1-\nu) q^{2\nu} }{\Gamma(\nu)} \frac{I_{-\nu}(q\ell)}{I_{\nu}(q\ell)}\right)|\a_q|^2+
		\right.
 \nonumber\\ 
& & \nonumber \\
		& &  + 
		\left( \frac{\Delta_-}{\ell^d} + \frac{(q\ell)I_{\nu-1}(q\ell)}{\ell^{d} I_{\nu}(q\ell)}\right)|\tphi_{\ell}(q)|^2
- \left. \frac{2^{1-\nu} q^{\nu}}{ \ell^{\frac{d}{2}}\Gamma(\nu) I_{\nu}(q\ell)} \left(\a_q^{\ast}\tphi_{\ell} + \tphi_{\ell}^{\ast}\a_q\right)\right\}
\label{eq:slactUV}
\end{eqnarray}
where we have assumed that $q$ is spacelike and have used  (\ref{eq:eomsol}) to express $\b_q$ in terms of $\alpha_q$ and $\tphi_\ell$.  For timelike momenta, we analytically continue $q = \sqrt{q^2} \to i |q|$, recalling that

\be
	J_{\nu}(|q|z) = e^{-i \pi \nu} I_{\nu} (i |q| z)
\ee


\subsubsection{Computing $\Psi_{IR}$}

The computation of $\Psi_{IR}(\phi_{\ell}) = e^{ i S_{IR}(\phi_{\ell})}$ requires specifying the behavior of $\phi$ as $z \to \infty$.  We will assume that in the Euclidean continuation, $\phi$ is nonsingular in this limit, as expected for computations in the vacuum state of the theory. The resulting solutions to the equations of motion with spacelike momenta and with the boundary conditions $\tphi(z=\ell,q) = \tphi_{\ell}$ are
\be	
	\tphi(z,q) = \left(\frac{z}{\ell}\right)^{\frac{d}{2}} \frac{K_{\nu}(qz)}{K_{\nu}(q\ell)} \,\tphi_{\ell}(q)\label{eq:irsoln}
\ee
For timelike momenta, we simply continue $q \to i |q|$.  Inserting (\ref{eq:irsoln}) into the supergravity action for $z > \ell$, we find that
\be
	S_{IR} = \frac{N^2}{2} \int \frac{d^d q}{(2\pi)^d} \left[ \frac{\Delta_-}{\ell^d} + \frac{(q\ell)K_{\nu-1}(q\ell)}{\ell^d K_{\nu}(q\ell)}  \right]|\tphi_{\ell}|^2 \label{eq:irone}
\ee
In Lorentzian signature, other boundary conditions at $z\to\infty$ are allowed.  

\subsection{Scheme dependence and the interpretation of $\Psi_{IR}$\label{schdep}}

Eq. (\ref{eq:hpansatz}) identifies $\Psi_{IR}$ with the generating function of correlators of single-trace operators in the cutoff theory. We can investigate this further by studying the two-point function in more detail, for $\nu \notin \Z$.
%
Using (\ref{eq:hpansatz}) and (\ref{eq:irone}), the  two-point function $\langle\O(q)\O(p)\rangle = \delta(q+p) G_{(2)}(q)$ in the cutoff theory is:
\be\label{eq:unsubgf}
	G_{(2)}(q;\ell) = \Delta_- \ell^{- 2\nu}  - (q\ell)\ell^{-2\nu} \frac{I_{-\nu + 1}(q\ell) - I_{\nu-1}(q\ell)}{I_{-\nu}(q\ell) - I_{\nu}(q\ell)}
\ee
Here we have written $K_{\nu}$ in terms of $I_{\pm\nu}$, for the case $\nu\notin \Z$.

In the bare theory defined at $z = \eps \to 0$, we expect $G_{(2)}(q) \sim q^{2\nu}$, by conformal invariance.  In practice, this scaling emerges after holographic renormalization to remove divergences and contact terms.  In the present theory, we expect this scaling to emerge when $q\ell \ll 1$. Using $I_{\pm\nu}(x) \sim x^{\pm\nu}(1 + \O(x^2))$, $G_{(2)}(q)$ can be seen to have a series of terms analytic in $(q\ell)^2$, as well as a series of non-analytic terms proportional to $q^{2\nu m + 2n}\ell^{2n}$, where $m > 0, n \geq 0$ are integers. To make this structure explicit, let us expand $G_{(2)}$ in the following fashion\footnote{In the range $2 > \nu > 1$, $I_{-\nu}(q\ell)$ has a zero for $(q\ell) = f(\nu) \sim \O(1)$; therefore, this expansion of $G_{(2)}$ is only valid for momenta smaller than $f(\nu)$.}
%
\begin{eqnarray}	
	G_{(2)}(q;\ell) & = & \left[ \Delta_- \ell^{- 2\nu}  - (q\ell)\ell^{-2\nu} \frac{I_{-\nu + 1}(q\ell)}{I_{-\nu}(q\ell)}\right] + \nonumber\\
	& & \ \ \ \ \ + \left[ \ell^{-2\nu} \frac{(q\ell)I_{\nu-1}(q\ell)}{I_{-\nu}(q\ell)} - \frac{(q\ell)I_{-\nu+1}(q\ell)I_{\nu}(q\ell)}{I_{-\nu}(q\ell)^2}\right] + \CO((q\ell)^{4\nu}) + \ldots \nonumber \\ \label{eq:tpdexpand}
\end{eqnarray}
All the terms on the first line are analytic in the momenta, whereas the terms on the second line are all nonanalytic for $\nu \neq \mathbb{Z}$.

The term proportional to $q^{2\nu}$ yields the expected behavior of the two-point function of the Fourier modes of an operator $\O$ with dimension $\Delta= \frac{d}{2} + \nu$.    The additional momentum dependence, subleading in $(q\ell)$, can occur if $\O$ mixes with $\p^k \O$, as one might expect from a theory with a cutoff.   The term scaling as $q^{4\nu + k}$ has the scaling expected from the mixing of $\CO$ with $\p^k \CO^2$, and so on.  Finally, any given term proportional to $(q\ell)^{2k}$ leads to a contact term in the two-point function, proportional to a $(2k)^{th}$ derivative of a delta function; this arises from mixing with the identity.\footnote{See Section 1 of \cite{Closset:2012vp}, especially the first two pages, for a clear general discussion of contact terms.}\footnote{Scheme dependence in holographic renormalization at the boundary $z \to 0$, with a similar discussion of operator mixing, has been discussed in \cite{Borodatchenkova:2008fw}.}  The infinite series of such terms sums up to a contribution to $G_{(2)}$ which is nonlocal on the scale of the cutoff. Note that if we take $\ell \to 0$, then all but a finite number of divergent, local terms,  vanish.  However, the cutoff theory is nonlocal at the scale $\ell$, and we do expect further terms analytic in $(q\ell)^2$.  

In summary, in the scheme used by \cite{Heemskerk:2010hk}, the field $\phi_{\ell}$ at the cutoff surface appears to couple to a mixture of the operator $\CO$ and other operators in theory with different scaling dimensions.   To deal with the mixing with the identity (that is, with the contact terms), we introduce a ``renormalized" $\Psi_{IR}$ via:

\be
\Psi_{IR}(\phi_{\ell}) = e^{i S_{ct}^{\ell}(\phi_\ell)} \, \Psi_{IR}^{ren}(\phi_{\ell}) = e^{i S_{ct}^{\ell}(\phi_\ell)} \langle e^{i N^2 \int d^d x \, \ell^{-\Delta_-} \phi_{\ell}\O} \rangle_{ren,\ell^{-1}}
\label{anotherscheme}
\ee
We will treat $\Psi_{IR}^{ren}(\phi_{\ell})$ as the generating function of correlators in the renormalized cutoff theory.   We will simultaneously multiply $\Psi_{UV}$ by $e^{i S_{ct}^{\ell}(\phi_\ell)}$ so that the complete theory is unchanged. 
With this renormalization of $\Psi_{IR}$, we can rewrite the supergravity path integral (\ref{eq:segmented}) as:

\be
	\mathcal{Z}[\a] = \int \D\phi_{\ell} \, \Psi_{UV}^{ren}(\phi_{\ell},\a)\, \Psi_{IR}^{ren}(\phi_{\ell}) 
\ee
where

\be
	\Psi_{UV}^{ren} = e^{i S_{UV}(\phi_{\ell}) + i S_{ct}^{\ell}(\phi_{\ell})} \equiv e^{i S_{UV}^{ren}(\phi_{\ell})}
\ee
We place the following requirements on $S_{ct}^{\ell}$:

\bi
\item $S_{ct}^{\ell}$ is constructed from $\phi_{\ell}(x)$ and its derivatives along the cutoff surface. This guarantees that it is analytic in the momenta and that its variation is consistent with Dirichlet boundary conditions at $z=\ell$.

\item An attractive feature is to require that the terms in $S_{ct}^{\ell}$ which diverge as $\ell \to 0$  agree with the usual UV counterterms $S_{ct}^\e$ as $\ell \to \eps$.  This ensures that $\Psi_{UV}^{ren}$ does not diverge as $\ell \to 0$, and that $\Psi_{IR}^{ren}$ becomes identical with conventional partition function of the full theory.\footnote{As $\eps \to 0$, this guarantees the the counterterm can be written as a total derivative in the bulk as in \cite{Dong:2012af}.}
\ei

\noindent The above proposal implies that the single- and multi-trace couplings will be derived from $S_{UV}^{ren}$.  In particular, if
\be
	S_{UV}^{ren} = \frac{N^2}{2} \int \frac{d^d q}{(2\pi)^d} \left\{ \frac{1}{\ell^{2\Delta_-} h(q)} \big|\tphi_{\ell}(q) - \ell^{\Delta_-} \lambda_q(\a_q)\big|^2 + C(q,\lambda_q)\right\}
\ee
then we can write

\be
	Z = \Big\langle e^{i N^2 \int \frac{d^dq}{(2\pi)^d} \left(\lambda_q {\tilde\O}(q) + \frac{1}{2} h(q) {\tilde\O}^2 + C(q,\lambda_q)\right)}\Big\rangle_{\ell}
\ee
In this equation $C(q)$ is the coupling to the cosmological constant. Note that while $S_{UV,IR}$ in the ``bare" theory (without counterterms) satisfied the Hamilton-Jacobi equations which follow from supergravity, the counterterms will in general modify these equations \cite{Dong:2012af}.

\subsubsection{A minimal subtraction scheme}

In this subsection, we will work with $d = 4$.  The simplest counterterm satisfying the criteria we have demanded is:
\be
	S_{ct}^{\ell} = \frac{N^2}{2} \int \frac{d^4 q}{(2\pi)^4} \, \left(\frac{\Delta_-}{\ell^4} + \frac{q^2}{\ell^2(2\nu-2)}\right) \tphi_{\ell}^2
\ee
The renormalized two-point function of single-trace operators in the cutoff CFT is:
\be
	G_{(2)}(q) = \frac{(q\ell)K_{\nu-1}(q\ell)}{\ell^d K_{\nu}(q\ell)} - \frac{q^2}{\ell^2(2\nu-2)}
\ee
For $\ell \to 0$ this will be proportional to $q^{2\nu}$.  For finite $(q\ell) < f(\nu)$ (where $f(\nu)$ is the smallest zero of $I_{-\nu}(x)$), there will be an infinite series of subleading contact terms, summing to a contribution that is nonlocal at the scale $\ell$.

Using this prescription, the renormalized UV action reads

\begin{eqnarray}
\hspace{-1cm}	S_{UV}^{ren} & = & \frac{1}{2} \int \frac{d^4 q}{(2\pi)^4} \left[ \left( \frac{(q\ell)I_{\nu-1}(q\ell)}{\ell^{4} I_{\nu}(q\ell)} - \frac{(q\ell)^2}{\ell^4 (2-2\nu)} \right) |\tphi_{\ell}|^2 + \right. 
	\nonumber\\
 && \nonumber\\	
	& &  + \frac{2^{1-2\nu} \Gamma(1-\nu) q^{2\nu} I_{-\nu}(q\ell)}{\Gamma(\nu) \, I_{\nu}(q\ell)} |\a_q|^2
 + \left. \frac{2^{1-\nu} q^{\nu}}{\ell^{2}  \, \Gamma(\nu) \, I_{\nu}(q\ell)} (\a_q^{\ast}\tphi_{\ell} + \a_q \tphi_{\ell}^{\ast}) \right]
\end{eqnarray}
The resulting single-trace coupling is

\be
	\lambda_q = \frac{2^{1-\nu}  q^{\nu}\ell^{2} \, \a_q}{\Gamma(\nu) \, \left[ (q\ell) I_{\nu-1}(q\ell) - \frac{(q\ell)^2 I_{\nu}(q\ell)}{(2-2\nu)} \right]} 
\ee
As $\ell \r 0$, the single trace coupling\footnote{The two-point correlation function in the full Wilsonian theory is computed in $\S 4.$}  has the expected scaling $\l_q \r \ell^{\Delta_-} \a_q$.  The double-trace coupling is:
\be\label{eq:gausstpfmin}
	h(q) = \frac{\ell^4 I_{\nu}(q\ell)}{(q\ell) I_{\nu-1}(q\ell) - \frac{(q\ell)^2 I_{\nu}(q\ell)}{2-2\nu}}
\ee
Note that this is nonlocal on the scale $\ell$ and is independent of $\alpha_q$.  In particular, it is nonzero even for an {\it unperturbed}\ conformal field theory.  This may seem surprising.  We will return to this point in \S4. 

\subsubsection{A maximal subtraction scheme}

When $q\ell < f(\nu)$, the part of $G_{(2)}$ analytic in $(q\ell)^2$ is contained in the first line of (\ref{eq:tpdexpand}), and represents an infinite series of contact terms in the two-point function.   For such sufficiently low momenta, we can completely remove the contact terms by adding the following IR counterterms

\be
	S_{ct}^{\ell} =  \frac{N^2}{2} \int \frac{d^d q}{(2\pi\ell)^d} \left(\Delta_- + (q\ell)\frac{I_{1-\nu}(q\ell)}{I_{-\nu}(q\ell)}\right)|\tphi_{\ell}|^2\label{eq:maxct}
\ee
With this choice, the renormalized two-point function for the cutoff theory is:

\be
G_2^{ren} = \frac{1}{I_{-\nu}(q\ell) K_{\nu}(q\ell)} \sim q^{2\nu} \left(1 + \frac{(q\ell)^2}{2(\nu-1)} + \ldots \right) + \O(q^{4\nu})\label{eq:rentp}
\ee
(Note that we have not removed mixing with $\p^k \O$ and with $\p^k \O^2$). The renormalized UV action $S_{UV}^{ren}$ now takes the simple form:

\be
	S_{UV}^{ren} = \frac{N^2}{2}  \int \frac{d^d q}{(2\pi)^d} \frac{2  \sin \pi\nu}{\pi \, \ell^d I_{\nu}(q\ell) I_{-\nu}(q\ell)} \Big| \tphi_{\ell}(q) - \ell^{\frac{d}{2}} I_{-\nu}(q\ell)\Gamma(1-\nu) (q/2)^{\nu} \a_q\Big|^2
\ee
After integrating over $\tphi_\ell$, this yields the following Wilsonian action

\be
	\mathcal{Z} = \Big\langle \exp \left( i N^2 \int d^d x\, \ell^{- \Delta_-} g_{ren}(x;\ell) \O(x)  + \frac{iN^2}{2} \int d^d x d^d y \, h_{ren}(x-y;\ell) \O(x) \O(y) \right)\Big\rangle
\ee
where

\be\label{eq:maxsst}
	{\tilde g}_{ren}(q;\ell) = \int d^d x \, e^{i q\cdot x} g_{ren}(x;\ell) = \ell^{\frac{d}{2}} \left(\frac{q}{2}\right)^{\nu} I_{-\nu}(q\ell) \Gamma(1-\nu) \a_q
\ee
and
\be\label{eq:gausstpfmax}
	{\tilde h}_{ren}(q;\ell) = \int d^d (\Delta x)\, h_{ren}(\Delta x;\ell)\, e^{i q\cdot \Delta x} = \frac{I_{\nu}(q\ell) I_{-\nu}(q\ell) \pi \ell^{-2\Delta}}{2 \sin\pi\nu}
\ee
Let us discuss each of these couplings in turn.

The scheme we have chosen, based on the desire to remove all contact terms from $G_{(2)}(q)$, has the interesting feature that the single-trace coupling $g_{ren}(x,\ell)$ satisfies the bulk equations of motion if we substitute $z$ for $\ell$.  This is evident from a comparison of (\ref{eq:maxsst})\ with (\ref{eq:eomsol}).  This is reminiscent of the previous interpretations of the running couplings in AdS/CFT, prior to \cite{Heemskerk:2010hk,Faulkner:2010jy}, in terms of solutions to the equations of motion.  However, note that the solution appearing here as the coupling, $z^{d/2} I_{-\nu}$, is different from the one employed in previous work, which was $z^{\frac{d}{2}} K_\nu$. In particular, it contains no subleading piece scaling as $z^{\Delta}$ -- i.e., the kind that is usually associated with the classical expectation value of the operator \cite{Balasubramanian:1998sn,Balasubramanian:1998de}. One may worry that, unlike $K_\nu$, the Bessel function $I_{-\nu}$ diverges as $q\ell \r \infty$. However, in the IR theory, we should only consider $q\ell \ll 1$, so this issue does not arise.

In a general scheme, $g_{ren}$ does not solve the equations of motion; in particular, the single-trace coupling found in \cite{Heemskerk:2010hk,Faulkner:2010jy}\ does not. In fact, we found that no addition of local counterterms to $S_{ct}^{\ell}$  will lead to a solution which is a general combination of the form (\ref{eq:eomsol}) with $\a_q \cdot \b_q \neq 0$.  This will become apparent following the general discussion in \S3.3.
For the double-trace couplings we find, as in the minimal subtraction scheme, that $h_{ren}(x-y;\ell)$ is generically nonlocal at the scale $\ell$, and induced even when the single-trace perturbations vanish. We will discuss and interpret this phenomenon in \S4. Note that in all schemes $h \propto \ell^{d}$ as $\ell \r 0$.

While this is an attractive scheme, it only makes sense when $q\ell$ is less than the smallest zero of $I_{-\nu}$.  This zero leads to a pole\footnote{ Note that there is no pole the 2-point function per se -- the singularity is in the expansion that separates the analytic and non-analytic pieces.} in the counterterm (\ref{eq:maxct}) and in the renormalized two-point function (\ref{eq:rentp}), when $\nu < 2$.  In general, we should not be worried about these divergences, since in a Wilsonian framework we should not be  discussing couplings to momentum modes with $q\ell \gtrsim 1$. Consequently, this scheme seems generally sensible. However, it fails is in the limit $\nu \to 1$ or $\nu \to 2$, at which point the zero\footnote{When $\nu = 2 -\delta$ with $\delta <<1$, the zero is at $q\ell \sim \delta^{\frac{1}{4}}$. Thus, $\nu$ has to be extremely close to an integer in order for the scheme to break down.} of $I_{-\nu}$ moves to $q\ell = 0$.  In this limit, the two-point function picks up a logarithmic term, and we we have not found a good analog of the ``maximal subtraction" scheme.   

\subsection{Maximal subtraction scheme  for interacting theories}

In the prior subsection, we found that for sufficiently small momenta, there exists a natural scheme for the Wilsonian action in which the single-trace couplings are proportional to the purely non-normalizable solution in (\ref{eq:eomsol}) - i.e, the solution with $\beta_q = 0$.  In this subsection, we will sketch an argument that such a scheme continues to exist in the interacting theory at large $N$, provided that the dual operator is relevant and  the interactions can be treated perturbatively.

\subsubsection{The free theory revisited}

To motivate this, let us provide a different construction of the ``maximal subtraction" counterterm in the free theory.  First recall from \S2.3\ that the single-trace coupling is equal to the value of $\phi_{\ell}$ which is a stationary point of $S_{UV}(\phi_{\ell})$.
Thus, our goal is to find a counterterm action $S_{ct}(\phi_{\ell})$ that is analytic in $\phi_{\ell}(x)$ and its derivatives, such that the family of extrema of $S_{UV}(\phi_{\ell}) + S_{ct,\eps} + S_{ct,\ell}(\phi_{\ell})$ treated as a function of $\ell$ matches the solution (\ref{eq:eomsol}) with $\beta_q = 0$.  Following the discussion in \S2.3, this will imply that the Wilsonian single-trace coupling as a function of $\ell$ will be (\ref{eq:eomsol}) with $\beta = 0$ and with $z$ replaced by $\ell$.

We start by noting that $S_{UV}(\phi_{\ell}) + S_{ct,\eps}$, as the classical action, satisfies the Hamilton-Jacobi equations for a free scalar field ($S_{ct,\eps}$ will not contribute at $z = \ell$).  A solution to the equations of motion, consistent with the boundary conditions at $z = \eps,\ell$, will then solve Hamilton's equation:
\be
	\frac{\delta {\cal H}}{\delta\pi} = \ell^d \frac{\delta S_{UV}(\phi_{\ell}(x))}{\delta \phi_{\ell}(x)}
		=  z\p_z \phi(z,x) \Big|_{z = \ell}\ ,
\ee
where we have used that fact that ${\cal H}$ is quadratic in $\pi$. We wish to find $S_{ct,\ell}$ such that
\be\label{eq:ctequation}
	\ell^d \frac{\delta S_{ct,\ell}(\phi_{\ell})}{\delta\phi_{\ell}} = - z\p_z \phi_n(z,x) \Big|_{z = \ell}
\ee
where $\phi_n$ is the solution to (\ref{eq:eomsol}) with $\beta_q = 0$, consistent with $\phi(\ell,x) = \phi_{\ell}(x)$.  It follows that
\be
	\ell^d \frac{\delta (S_{UV} + S_{ct,\ell})}{\delta \phi_{\ell}} = z\p_z \left(\phi(z,x) - \phi_n(z,x)\right)\Big|_{z = \ell}\ .
\ee
This will vanish when $\phi(z,x) = \phi_n(z,x)$, so that $S_{UV} + S_{ct,\ell}$ is stationary when $\phi_{\ell} = \phi_n(\ell,x)$.  Thus, $\phi_n(\ell,x)$ will be the single-trace coupling, following the logic discussed in \S2.3.

Let us now find the solution to (\ref{eq:ctequation}).  Taking the Fourier transform of $\phi_n(z,x)$:
\be\label{eq:nns}
	\tphi_n(z,q) = \alpha_q \Gamma(1-\nu) \left(\frac{q}{2}\right)^{\nu} z^{\frac{d}{2}} I_{-\nu}(qz)
\ee
we find that
\be
	z\p_z \tphi_n(z,q)\Big|_{z=\ell} = \left( \Delta_- + \frac{(q\ell)I_{1-\nu}(q\ell)}{I_{-\nu}(q\ell)} \right)\tphi_{\ell}\ .
\ee
Inserting this into (\ref{eq:ctequation}) and integrating leads to the counterterm of the ``maximal subtraction" scheme we have discussed in the previous section, up to a constant.\footnote{In the free theory, this constant can be shown to be zero. In the maximal subtraction scheme, $S_{IR}^{ren}(\phi_\ell)$ vanishes when evaluated on the purely non-normalizable solution. Since $S_{UV}^{ren} = S_{IR}^{ren}(\phi_\e)-S_{IR}^{ren}(\phi_\ell)$, it follows that $S_{UV}^{ren}$ also has  a zero when $\phi_\ell = \phi_n(\ell)$. } Note that $S_{ct,\ell}$ is Gaussian.  Thus $S_{UV,ren}$ is Gaussian in this free theory, with a minimum at $\phi_{\ell}(x) = \phi_n(\ell,x)$.

For $q\ell$ less than the smallest zero of $I_{-\nu}$, $S_{ct,\ell}$ is an analytic function of $(q\ell)^2$.  If we were to repeat the above procedure for $\phi_n$ of the form (\ref{eq:eomsol}) with $\alpha,\beta \neq 0$, we would find that $S_{ct}$ was no longer analytic; if we follow our criteria, we cannot find a scheme for which the single-trace coupling is a general solution to the equations of motion.  We have not, however, studied the effects of taking mixing of $\O$ with $\p^k \O$, $\O^2$, and so forth, into account.

\subsubsection{The interacting case}

Next we consider the interacting theory (\ref{eq:scact}), in the case that $V$ is polynomial in $\phi$ and the interactions can be treated perturbatively. We can then construct, in perturbation theory, a solution $\phi_n$ which is (\ref{eq:nns}) to zeroth order in $V$. If $V$ is a local functional of $\phi$, then the corrections to the leading order solution will be local powers of $\phi_n$ convolved against products of the Green function for the free theory. Now, in momentum space along the boundary, the kinetic term depends on $q$ as $(qz)^2$.  The Green function, the inverse of this operator, will have Dirichlet boundary conditions at $z = \eps,\ell$.  Thus it can be expanded in a power series in $(qz)^2$ with integer powers.  Similarly, we can write $z\p_z \phi_n(z,x)\Big|_{z = \ell}$ as a power series in $\phi_n(\ell,z)$ and $(q\ell)^2$.\footnote{We have not checked the radius of convergence of this power series.}

Following the procedure above, we define $S_{ct,\ell}$ as the solution to the functional differential equation
\be
	\ell^d \frac{\delta S_{ct,\ell}}{\delta \phi_{\ell}} = - z\p_z \phi_n(x,\ell)\Big|_{z = \ell}
\ee
where $\phi_{\ell}(x) = \tphi_n(\ell,x)$.  Modulo issues of the radius of convergence, we can expand the right hand side in a power series in $\phi_{\ell}$ and in $(q\ell)^2$. Once we have found this, we again find that
\be
	\ell^d \frac{\delta (S_{UV} + S_{ct,\ell})(\phi_{\ell})}{\delta \phi_{\ell}} = \ell^d \frac{\delta S_{UV,ren}(\phi_{\ell})}{\delta \phi_{\ell}} = z\p_z (\phi(z,x) - \phi_n(z,x))\Big|_{z = \ell}
\ee
and thus $S_{UV,ren}$ is stationary when $\phi(z,x) = \phi_n(z,x)$. The result, following the logic of \S2.3, is that the running single-trace coupling will be $\phi_n(\ell,x)$.

We conclude by noting that as $\ell \to 0$, $S_{ct,\ell}$ will approach $S_{ct,\eps}$. For a general solution to the equations of motion, the divergences in $S_{UV}$ are known to be equivalent to those which would arise if the solution were $\phi_n$ \cite{Skenderis:2002wp}.  Thus we can use the same procedure to construct $S_{ct,\eps}$ unambiguously in the limit $\eps \to 0$, and it is guaranteed to be equivalent to $S_{ct,\ell}$ as $\ell \to \eps$.

\section{Multiple-trace operators and UV/IR entanglement}

\def\tlh{{\tilde h}}

In this section we investigate the double-trace operators induced in the prescription of \cite{Heemskerk:2010hk,Faulkner:2010jy}. Specifically, we would like to address two facts about the double-trace operators we computed in \S3.1: they are nonlocal on the scale of the cutoff, and they are induced for {\it all}\ single trace operators $\O$, whether or not the UV theory is perturbed, and whether or not $\O$ is relevant or irrelevant.

In \S4.1\ we will discuss the bulk interpretation of these double-trace operators. We will show that by first computing $S_{UV}$, and considering it as a boundary term for the IR region, the result is a theory in the IR region with mixed Dirichlet-Neumann  boundary conditions at $z = \ell$ on classical solutions (see also the discussion in \cite{Faulkner:2010jy}). These boundary conditions encode the propagation of signals out of the IR region, through the UV region, and back. This provides another perspective on the emergence of double-trace operators.  Following this, in \S4.2\ we will discuss the spacetime interpretation of the multiple-trace operators in the free theory. 

\subsection{$\Psi_{UV}$ and the variational principle at the cutoff surface}

Let us start with the full supergravity path integral written in the form (\ref{eq:segmented}), and perform the path integral over the UV region $0 < z < \ell$.  The resulting expression is:
\be
Z = \int \D \phi_{\ell}\,  \D \phi_{z>\ell} \, e^{i S_{bulk}(z>\ell) + i S_{UV}(\phi_{\ell})} \label{eq:pathint}\ ,
\ee
and can be viewed as a path integral over the IR region with an additional boundary term provided by $S_{UV}(\phi_{\ell})$.
As we have already discussed, in the free theory $\Psi_{UV}$ is a Gaussian functional of $\phi_{\ell}$:
\be\label{eq:suvbound}
S_{UV} = -i \ln \Psi_{UV} = \frac{N^2}{2} \int \frac{d^d q}{(2\pi\ell)^d}\left( \frac{1}{\tilde h} \big|\tphi_{\ell} - \tilde \l\big|^2 + c\big|\tilde \l\big|^2\right)\ ,
\ee
in which
\be
\tilde h = \left( \Delta_- + q \ell \frac{I_{\nu-1}}{I_\nu}\right)^{-1} \;, \;\;\;\;\; \tilde \l = \frac{2 \left(\frac{q}{2}\right)^{\nu} \ell^{\frac{d}{2}} \a }{\Gamma(\nu) \left(I_\nu \Delta_- + q \ell I_{\nu-1}\right)} \ .
\ee
Here $c|{\tilde \lambda}|^2$ is a cosmological constant term, and the Bessel functions have the argument $(q\ell)$. If we demand that the path integral be stationary, we will find that the standard bulk equations of motion will emerge, along with the boundary equation of motion
\be\label{eq:boundeom}
z \p_z \tphi(z,q)\Big|_{z=\ell} - {\tilde h}^{-1}(q) \left(\tphi_{\ell}(q) - {\tilde \lambda} (q)\right) = 0\ .
\ee
These are mixed boundary conditions for classical solutions.  In general such boundary conditions imply that conservation of energy and probability are violated at $z = \ell$. This is unsurprising, since one expects many physical excitations in the IR regime to propagate to the UV region (small $z$). We will discuss this further in \S4.2.

We can rederive (\ref{eq:boundeom}) in the following way. We wish to demand that the boundary values of $\phi, \p_z\phi$ at $z = \ell$ are consistent with the boundary conditions $\phi(z,x) \to z^{d-\Delta} \alpha(x)$ as $z \to 0$.  If we fix the UV coupling $\a_q$, one can easily show - by eliminating $\b_q$ in the explicit solution (\ref{eq:eomsol})- that $\tphi_{\ell}$ and $\ell \p_\ell \tphi_\ell$ have to be related via  (\ref{eq:boundeom}).
Thus, the boundary conditions (\ref{eq:boundeom}) encode the classical propagation in the UV region consistent with Dirichlet conditions at infinity from the IR point of view.  In other words, the upshot is that $\Psi_{UV}$ transports the variational principle in the UV to a variational principle at the cutoff surface which is consistent with the equations of motion in the entire spacetime. The boundary conditions (\ref{eq:boundeom}) encode the classical propagation in the UV region consistent with Dirichlet conditions at infinity.

When $\nu < 1$, mixed boundary conditions near the UV boundary are known to describe perturbations by multitrace operators. Here, $\l$ corresponds to the coupling to the single-trace operator, and $h$ to the coupling to the double trace operator.\footnote{The usual discussion is in terms of the alternate quantization of \cite{Klebanov:1999tb}, since then the double-trace perturbation is relevant.  The standard quantization, in which the perturbation is irrelevant, appears as the IR fixed point of this theory \cite{Witten:2001ua}.  The flow close to this fixed point is controlled by the double-trace operator in the standard quantization, which corresponds to $1/\gamma$.} We would like to argue for all $\nu$, at finite cutoff, the running double-trace operator is similarly associated to mixed boundary conditions.

In the large $N$ limit, using a Hubbard-Stratonovich transformation in the field theory path integral, one can show that the two-point function $G_h$ of a single-trace operator $\O$ in the presence of a double-trace perturbation $h \O^2$ is given by (see for example \cite{Faulkner:2010jy,Mueck:2002gm,Hartman:2006dy}):
\be
G_h = \frac{G}{1+hG} \label{eq:hstrat}
\ee
where $G$ is the two-point function for $\O$ in the large-$N$ theory in the absence of a double trace perturbation.  

In the full AdS/CFT correspondence, the same relationship is present between the holographic two-point functions in the presence of Dirichlet vs mixed boundary conditions \cite{Mueck:2002gm,Hartman:2006dy}. In this sense, mixed boundary conditions are directly related to double-trace operators. In our case, take $G_{(2)}$ to be the two-point function computed by differentiating the cutoff partition function $e^{i S_{IR}(\phi_\ell)}$ with respect to $\phi_{\ell}$, and take $G_{(2)}^{h}$ to be the Wilsonian two-point function computed by differentiating the full supergravity action, minus the cosmological constant, with respect to $\lambda$. The relationship between $G_{(2)}$ and $G_h$ is precisely the same as in \eqref{eq:hstrat}. This provides another justification for the association of the Wilsonian double-trace coupling with mixed boundary conditions.

We can now consider the change of renormalization scheme in this context.  If we subtract counterterms from the IR action, shifting $G \to G_{ren}$, the full action does not change.  The renormalized $S_{UV}$ takes the form (\ref{eq:suvbound}) with $(h, \lambda,c)$ replaced by $(h_{ren}, \lambda_{ren}, c_{ren})$. The same argument as the previous paragraph goes through, with the $G, G_h$ replaced the corresponding renormalized two-point functions

\FIGURE{
\includegraphics[scale=.55]{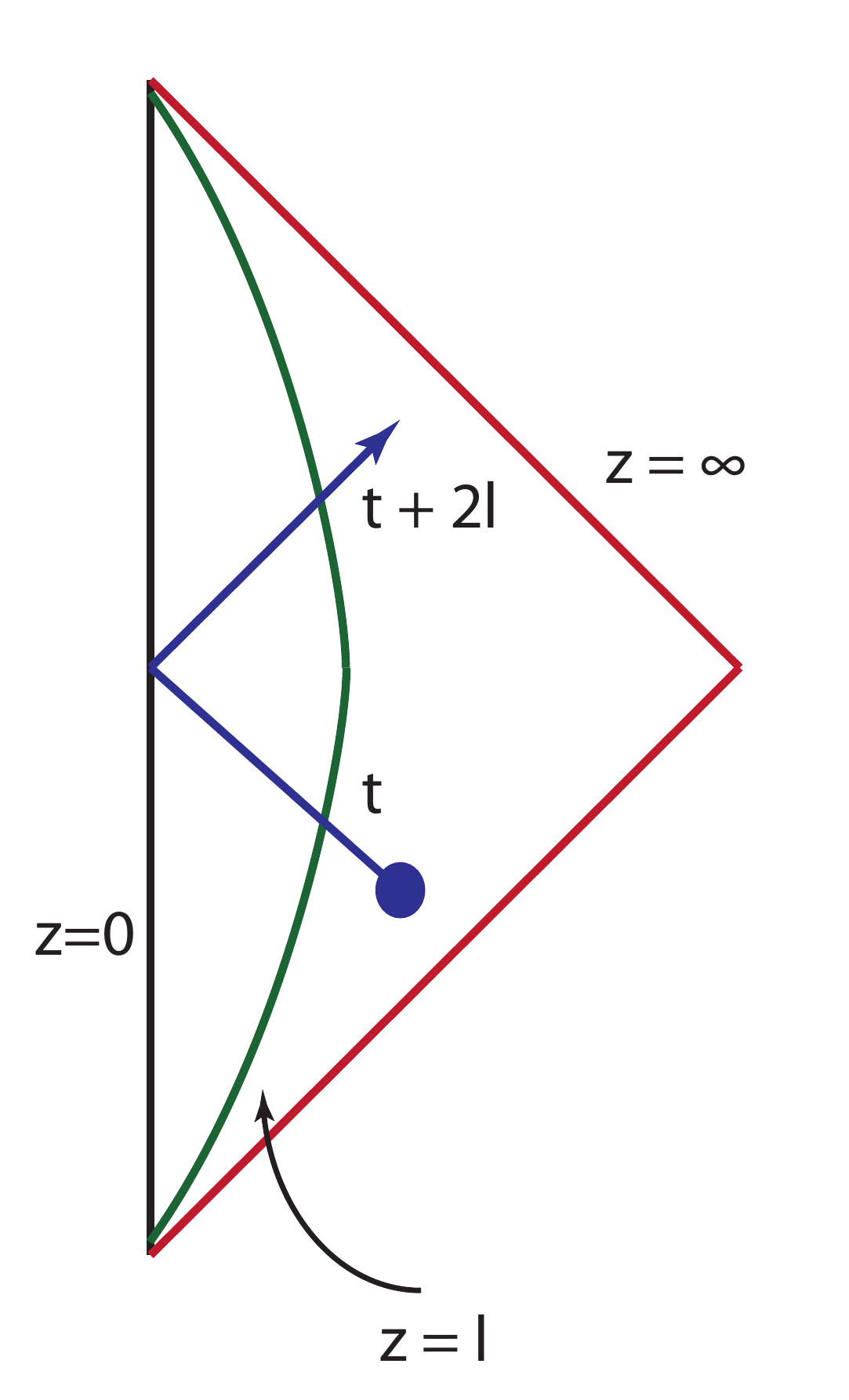}
\caption{A massless particle in the Poincar\'e patch, traveling from the IR region $z > \ell$ to the UV region $z < \ell$, bouncing off the $z = 0$ boundary, and continuing back into the IR region. The crossing times at the $z = \ell$ cutoff are $t, t + 2\ell$.}
\label{doubletrace}
}

\subsection{Double-trace operators in free supergravity}

In interacting quantum field theories, different spacetime scales are coupled, and excitations propagate across scales. In this subsection, we argue that the multiple-trace operators, which are induced in the framework of \cite{Heemskerk:2010hk,Faulkner:2010jy} even for an unperturbed CFT, serve to encode the propagation of information from the IR to the UV and back. 

Let us first consider the free bulk theory. Consider a holographic cutoff at $z = \ell$, and a massless scalar field $\phi$ dual to a marginal operator $\O$.  A scalar particle emitted from the $z > \ell$ region, at low enough momentum, will be dual to a mode of $\O$ acting on the vacuum at scales below the cutoff.  In the bulk, the particle will travel along the light cone towards either the boundary of AdS or the Poincar\'e horizon.  In the former case, as shown in Figure [\ref{doubletrace}], excitations bounce off of the boundary in a finite AdS time, and return to the origin.

If the massless particle first crosses $z = \ell$ at time $t$, it will cross back to the IR region at time $t' = t + 2 \ell$.  From the point of view of the cut off field theory, it is as if the excitation was absorbed at $t$ and re-emitted at $t + 2\ell$.  Following \cite{Balasubramanian:1998de,Banks:1998dd}, the creation and annihilation operators for $\phi$ are dual to momentum modes of $\O$ with negative and positive frequencies. Thus, the process would be described from the point of view of an IR observer as being generated by a double-trace perturbation of the field theory which is nonlocal in space and time, of the form
\be\label{eq:dt}
	\delta S = \frac{N^2}{2} \int d^d x\, d^d y\, h_{\ell}(x-y) \O(x)\O(y)\ ,
\ee
where $h_{\ell}$ is a nonlocal kernel smeared out over the scale $\ell$.  The double-trace couplings (\ref{eq:gausstpfmin},\ref{eq:gausstpfmax}) are of precisely this form. This suggests that the spacetime interpretation of (\ref{eq:dt}) is precisely that it describes the phenomenon illustrated in Fig. (\ref{doubletrace}). 

This phenomenon will occur in pure $AdS$ spacetimes dual to conformal field theories, as well as in duals to renormalization group flows between different strongly coupled fixed points.  More generally, operators of the form (\ref{eq:dt}) will be induced in the Wilsonian renormalization of any strongly interacting theory, in which degrees of freedom at different scales are coupled.  

Operators such as (\ref{eq:dt}) do not have an interpretation as perturbations of a Hamiltonian for the IR degrees of freedom, as they are nonlocal in time as well as in space. The Fourier transform of $h_{\ell}$ can be expanded in a power series in $(q\ell)$:
\be	
	\tlh_{\ell}(q\ell) = \sum_{n=0}^{\infty} \tlh_n (q\ell)^{2n}
\ee
We can thus write 
\be
	h_{\ell}(x-y) = \sum_{n = 0}^{\infty} \tlh_n (- \ell^2 \p_x^2)^n \delta(x-y)
\ee
In other words, $\delta S$ can be expressed as a sum over local higher-derivative operators.  This will include operators with arbitrarily high time derivatives.  

If we compute time-dependent correlation functions of IR operators by first tracing out the UV modes, the state of the IR modes would be described by a density matrix, as the UV and IR modes are highly entangled even in the ground state \cite{Balasubramanian:2011wt}.\footnote{We have not dealt with this issue here.  Rather we are studying correlators in the vacuum of the full theory, expressed in terms of infrared data.}  The higher-derivative operators contain information about the evolution of the density matrix.  In the theory of open quantum systems, this information is specified in addition to the Hamiltonian for the infrared degrees of freedom.

However, if we consider correlation functions of operators smeared out over distances $L\gg \ell$, the terms $h_n$ will give contributions suppressed by a factor $(\ell/L)^{2n}$ relative to the local $h_0 \O^2(x)$ term. If $\O$ itself is a good local operator describing a deformation of the Hamiltonian, the $h_0$ term can also be considered a local perturbation of the Hamiltonian; the effects of information passing into the UV degrees of freedom for some time will be seen only at scales $L \leq \ell$.

\FIGURE{
\includegraphics[scale=.5]{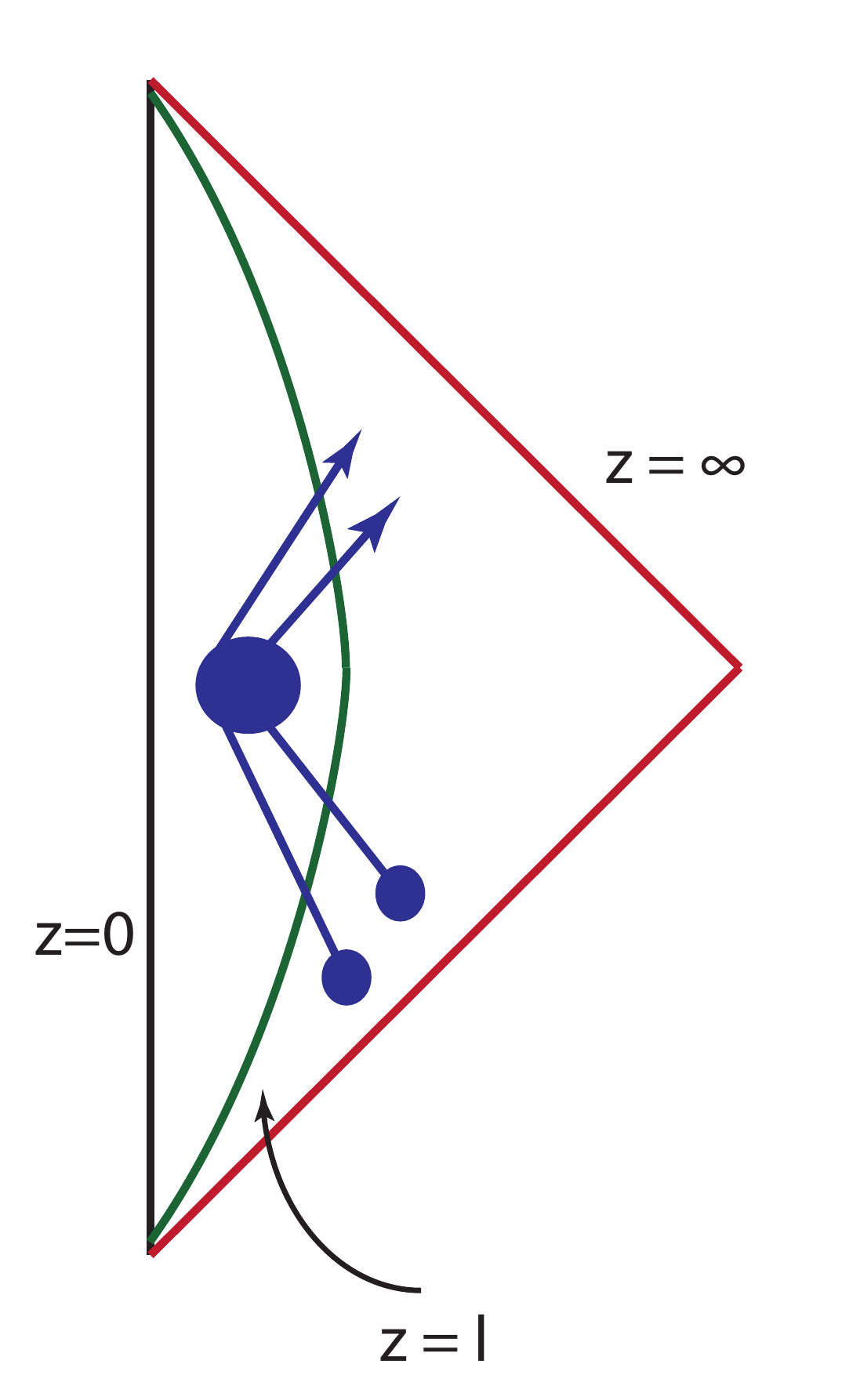}
\caption{Two particles propagate into the UV region, scatter via a tree-level quartic interaction, and propagate back into the IR region.}
\label{fourtrace}
}

When the bulk theory is interacting. triple- and higher-trace operators are generically expected to appear at leading order in the $1/N$ expansion, as described in \S2.3.   We can understand the multiple-trace terms along the same lines as the double-trace terms.  Consider two particles, modes of a single-trace operator $\O$, traveling from the IR to the UV region.  These particles can scatter in the UV region before returning to the IR region as shown in Figure (\ref{fourtrace}).  From the point of view of the IR variables, the connected part of this process will be described by a 4-trace interaction of the form
\be
	\delta S_{4} = \int d^dx_1\ldots d^d x_4 h_{\ell}^{(4)}(x_1,\ldots, x_4) \O(x_1)\ldots \O_4(x_4)
\ee

\section{Conclusions}

We can see a number of directions in which this formalism should be further developed.  We name two, related to physics in Lorentzian signature.

First, we recalled in \S2\ that one problem with identifying the bulk scalar fields as running couplings is that in Lorentzian signature, the value of the scalar in the bulk is affected by both the bare coupling and the choice of state.  Nonetheless, as in \cite{Balasubramanian:1999jd}, \cite{Heemskerk:2010hk,Faulkner:2010jy}\ do identify the value of the scalar field at the cutoff with the single-trace coupling in the {\it cut off}\ CFT.  This identification should be tested more and understood better. More generally, it is unclear to what degree the path integrals defining $\Psi_{UV,IR}$ can be made sensible in a full theory of quantum gravity.

Secondly, the calculations in this paper can be interpreted as calculating vacuum correlation functions in the full theory, in which one first does a path integral over the UV modes.  However, because the theory is strongly coupled, the infrared and ultraviolet degrees of freedom are strongly entangled \cite{Balasubramanian:2011wt}.  This means that the infrared degrees of freedom are properly described via a density matrix.  The evolution of this density matrix could be described in path integral language in terms of a Feynman-Vernon influence functional \cite{Feynman:1963fq}. This picture is, to our knowledge, undeveloped even in weakly-coupled quantum field theory.  At strong coupling, it would be very interesting to find the dual of this density matrix.

\vspace{0.25 in}

\paragraph{Acknowledgments: }

 We thank Tom Faulkner, Matthew Headrick, Suvrat Raju,  Eva Silverstein,  Balt van Rees and Herman Verlinde for helpful conversations.   This work was initiated at the Kavli Institute for Theoretical Physics (supported by the NSF grant  PHY-1125915) and continued at the Aspen Center for Physics (supported by NSF grant 1066293). AL is supported by DOE Grant DE-FG02-92ER40706.   VB and MG are supported by DOE grant DE-FG02-05ER- 41367.

\eject
\bibliographystyle{utphys}
\bibliography{hrgrefs}

\end{document}